\documentclass[draftclsnofoot, onecolumn, 12pt]{IEEEtran} % 单栏双行距草稿格式
\IEEEoverridecommandlockouts
\usepackage{cite}
\usepackage{amsmath,amssymb,amsfonts,comment,lipsum}
\usepackage{algorithmic}
\usepackage{graphicx}
\usepackage{textcomp}
\usepackage{xcolor}
\usepackage{stfloats}
\usepackage{float}
\usepackage{stfloats}
\usepackage{subfig}
\usepackage{multicol}
\usepackage{setspace}
 \usepackage{multirow}
\usepackage{hyperref}
\usepackage{algorithm}
\usepackage{setspace} % 设置双行距

\def\BibTeX{{\rm B\kern-.05em{\sc i\kern-.025em b}\kern-.08em
    T\kern-.1667em\lower.7ex\hbox{E}\kern-.125emX}}

\begin{document}

\title{Efficient Resource Allocation for Multi-User and Multi-Target MIMO-OFDM Underwater ISAC%
	\\
%{\footnotesize \textsuperscript	{*}Note: Sub-titles are not captured in Xplore and should not be used}
%\thanks{Identify applicable funding agency here. If none, delete this.}
}

%\vspace{-188 mm}

\author{
\IEEEauthorblockN{
Wei Men\textsuperscript{1,2,3§},
Longfei Zhao\textsuperscript{3},
Yong Liang Guan\textsuperscript{4},
Xiangwang Hou\textsuperscript{3},
Yong Ren\textsuperscript{3},
and Dusit Niyato\textsuperscript{5}
}

\IEEEauthorblockA{
\textsuperscript{1}\textit{National Key Laboratory of Underwater Acoustic Technology, Harbin Engineering University, Harbin 150001, China}
}

\IEEEauthorblockA{
\textsuperscript{2}\textit{College of Underwater Acoustic Engineering, Harbin Engineering University, Harbin 150001, China}
}

\IEEEauthorblockA{
\textsuperscript{3}\textit{Department of Electronic Engineering, Tsinghua University, Beijing 100084, China}
}

\IEEEauthorblockA{
\textsuperscript{4}\textit{School of Electrical and Electronic Engineering, Nanyang Technological University, Nanyang Avenue, 639798, Singapore}
}

\IEEEauthorblockA{
\textsuperscript{5}\textit{College of Computing and Data Science, Nanyang Technological University, Nanyang Avenue, 639798, Singapore}
}

\IEEEauthorblockA{
Emails: \{menwei, zhaolf23\}@tsinghua.edu.cn, eylguan@ntu.edu.sg, \{hxw21, reny\}@tsinghua.edu.cn, dniyato@ntu.edu.sg
}
 \thanks{\textsuperscript{§}Corresponding author: Wei Men (menwei@tsinghua.edu.cn).}
}

% \\ \vspace{-1.1em}\IEEEauthorblockA{  } 

\maketitle

\begin{center}
    \fbox{\parbox{\dimexpr\linewidth-2\fboxsep-2\fboxrule\relax}{\centering
    This work has been accepted for publication at IEEE GLOBECOM 2025.}}
\end{center}

\begin{abstract}
Integrated sensing and communication (ISAC) technology is crucial for next-generation underwater networks. However, covering multiple users and targets and balancing sensing and communication performance in complex underwater acoustic (UWA) environments remains challenging. This paper proposes an interleaved orthogonal frequency division multiplexing-based MIMO UWA-ISAC system, which employs a horizontal array to simultaneously transmit adaptive waveforms for downlink multi-user communication and omnidirectional target sensing. A multi-objective optimization framework is formulated to maximize the product of communication rate and range (PRR) while ensuring sensing performance and peak-to-average power ratio (PAPR) constraints. To solve this mixed-integer nonconvex problem, a two-dimensional grouped random search algorithm is developed, efficiently exploring subcarrier interleaved patterns and resource allocation schemes. Numerical simulations under real-world UWA channels demonstrate the designed system’s superiority and effectiveness: our algorithm achieves 90\% faster convergence than conventional exhaustive search with only a marginal 0.5 kbps$\cdot$km PRR degradation. Furthermore, the proposed resource allocation scheme maintains robustness beyond the baseline allocation schemes under stringent PRR and PAPR constraints.
\end{abstract}

%\addtolength{\topmargin}{-.48cm}
%\begin{IEEEkeywords}
%underwater acoustic sensor networks, integrated sensing and communication, waveform design and processing, multi-access interference suppression
%\end{IEEEkeywords}

\section{Introduction}
Internet of Underwater Things (IoUT) will play a key role in future ocean development\cite{11044529}. Exploiting underwater acoustic integrated sensing and communication (UWA-ISAC) technology has been considered an emerging solution that can greatly improve the efficiency of information collection and exchange of IoUT because it can share hardware, frequency, power, and other resources while avoiding mutual interference between sensing and communication functions. However, unlike simple point-to-point transmission and detection scenarios, UWA-ISAC applied to IoUT needs to serve multiple users simultaneously and detect potential targets in unknown directions. This seriously challenges integrated waveform design and the balance between dual functions.

A widely studied integrated waveform design philosophy is sensing-centric, embedding communication data into the sensing waveform, such as linear frequency modulation and generalized sine frequency modulation \cite{n6}. Unfortunately, this approach is challenging in meeting the needs of most communication scenarios due to the low data rate. Another alternative design philosophy is to focus on communication, that is, to directly use standard communication signals to perform target sensing. Orthogonal frequency division multiplexing (OFDM) has become one of the representative signals used in this method due to its high spectral efficiency and anti-multipath fading capabilities. More importantly, its range and velocity estimation performance have been powerfully demonstrated through theory and experiments \cite{n2,n3}. Consequently, OFDM-based ISAC has been widely studied in terrestrial and underwater applications. An OFDM pilot optimization scheme was studied in\cite{n4} to achieve a lower communication bit error rate while limiting the range error. In addition, OFDM can achieve lower range sidelobes through specific constellation mapping \cite{n2}, which makes it capable of scenarios in which multiple targets need to be detected.

Furthermore, to achieve higher sensing resolution and additional spatial diversity gain, recent work has studied ISAC technology based on multiple-input multiple-output (MIMO)-OFDM for independent execution of multi-user communication and multi-target sensing by combining beamforming technology \cite{n5}. Although such technology is urgently needed for IoUT with a large number of nodes, unlike RF channels, sea waves can cause the motion of the platform and underwater acoustic channels to vary over time, which results in prior information about communication users and potential targets of interest being partially or entirely missing. Therefore, it can be more efficient for all MIMO array elements to transmit orthogonal waveforms to each other to form a non-directional beam to cover communication users and targets. In \cite{n7,n9}, interleaved OFDM is adopted by MIMO radar/sonar to realize ISAC in arbitrary spatial domain, which has a higher spatial resolution in sensing and stronger robustness in communication.

Considering the resource-constrained nature of IoUT nodes, achieving optimal sensing and communication performance with limited transmission resources (e.g., frequency and power) remains a critical yet underexplored challenge for UWA-ISAC systems. To address this challenge, this paper extends our previous work \cite{n9} by designing a multi-objective UWA-ISAC framework, which can simultaneously optimize both communication and sensing by a novel resource allocation scheme that diverges from conventional approaches that separately optimize communication or sensing operations. Furthermore, we develop a two-dimensional grouped random search (TDGRS) algorithm to solve this nonconvex complex optimization problem efficiently.

\IEEEpubidadjcol
\textit{Notations:} 
Bold uppercase and lowercase letters represent matrices and vectors, respectively.
The notations $(\cdot)^{H}$, $(\cdot)^{T}$, ${{\left\| \cdot \right\|}_0}$ and ${{\left\| \cdot \right\|}_1}$ denote the Hermitian transpose, transpose, $\ell_0$ and $\ell_1$ norm, respectively. The notation $\mathrm{diag}(\cdot)$ forms a square matrix with the operand vector only on the main diagonal, and  $\operatorname{E} \left\{ \cdot \right\}$ denotes the expectation operator. Additionally, $\mathrm{Re} \left \{\cdot\right \}$ is the operator used to take the real part of the complex number.

\section{ MIMO-OFDM underwater ISAC signal model}

\subsection{System Model}
In this paper, we consider a UWA-ISAC scenario comprising a surface control node (SCN), $N_u$ seabed observation nodes (ONs), and $Q$ targets of interest, where the SCN enables simultaneous downlink multi-user communications to ONs and sensing of spatially distributed targets via its transmitted ISAC waveforms. Furthermore, during the pulse intervals of ISAC, communication users periodically transmit uplink OFDM signals to perform channel measurements and enable essential information exchange. In particular, the SCN is equipped with a horizontal monostatic sonar array with $M$ array elements arranged at equal spacing $d$. During operation, $M_t$ array elements are utilized for transmitting ISAC signals while $M_r$ array elements are dedicated to receiving uplink communication signals and target echoes, and neither $M_t$ nor $M_r$ is greater than $M$. For convenience, the sets of transmitting elements, receiving elements, and communicated ONs are indexed by ${\mathcal{M}_t} = \left[ {1,2, \cdots, {M_t}} \right]$, ${\mathcal{M}_r} = \left[ {1,2, \cdots, {M_r}} \right]$, and ${\mathcal{N}_u} = \left[ {1,2, \cdots, {N_u}} \right]$, respectively. In addition, all ONs use a single transducer for uplink signal transmission and ISAC signal reception, and transmission and reception take place in different time slots without interference.

Before transmitting the SCN's integrated waveform, multiple communication users send uplink pilot signals for channel measurement. Based on the measured channel state information (CSI), we generate adaptive downlink MIMO-OFDM waveforms with optimal resource allocation. Specifically, we divide an uplink-downlink cycle into two stages. Stage I: All $N_u$ ONs utilize the full frequency bandwidth to transmit $P$ pilot symbols for the SCN to perform channel estimation. Stage II: The SCN embeds $N_u$ independent data streams into the ISAC waveform for downlink transmission to communication users while exploiting the waveform for target detection.

\begin{comment}
\begin{itemize}
	
	\item Stage I: All $N_u$ communication users (i.e., ONs) utilize the full frequency  bandwidth to transmit $P$ pilot symbols for the SN to perform channel estimation.
	
	\item Stage II: The SN embeds $N_u$ independent data streams into the ISAC waveform for downlink transmission to communication users, while simultaneously exploiting the waveform for target detection.
	
\end{itemize}
\end{comment}

\vspace{-2mm}
\subsection{UWA Channel Model}
\vspace{-1.5mm}
The UWA channel has typical frequency-selective fading characteristics and it is usually time-varying due to transmitter-receiver motion. Assuming that the UWA channel between the SCN, and the $n$th communication user has $L_n^c$ paths, the frequency domain transfer function of the UWA channel at time $t$ is expressed as
\vspace{-2mm}
\begin{equation}
{H_n}\left( {f,t} \right) = \sum\limits_{l = 0}^{L_n^c - 1} {{A_{n,l}}\left( f \right){\beta _{n,l}}\left( {f,t} \right){e^{ - j2\pi f{\tau _l}\left( t \right)}}} , 
\end{equation}
where ${\tau _l}\left( t \right) = {\tau _l} - {\lambda _l} t $ represents the time-varying path delay, ${\lambda _l}$ denotes the Doppler factor of the $l$th path. ${\beta _{n,l}}\left( {f,t} \right)$ is the reflection coefficient of the $l$th path, and is normally equal to 1 for direct paths without bottom or surface reflections \cite{n3}. ${A_{n,l}}\left( f \right)$ denotes the gain of the $l$th propagation path, which is closely related to the transmission loss and the transmit-receive gain. According to the Urick model, the acoustic transmission loss includes spreading loss and absorption loss, which can be modeled as
\begin{equation}
{L_{n,l}}\left( {{r_{n,l}},f} \right) = {L_0}r_{n,l}^s\alpha {\left( f \right)^{\frac{{{r_{n,l}}}}{{1000}}}}, 
\end{equation}
where ${L_0}$ is a constant, $s$ is the spreading factor, $r_{n,l}$ is the path length, $\alpha \left( f \right)$ is the acoustic absorption coefficient of the seawater medium \cite{n3}. Thus, the path gain ${A_{n,l}}\left( f \right)$ can be further expressed as
\begin{equation}
{A_{n,l}}\left( f \right) = \frac{{{G_T}{G_R}}}{{\sqrt {{L_{n,l}}\left( {{r_{n,p}},f} \right)} }}, 
\end{equation}
where $G_T$ and $G_R$ are the transmit and receive gains, respectively.
Since both the SCN and ONs remain stationary or slow-moving in the considered ISAC scenario, the Doppler effect in the communication links is negligible, and the UWA channel is described as a quasi-static channel. In other words, we consider the CSI to be approximately time-invariant within a channel coherence period. Therefore, the frequency response of the UWA channel between the SCN and the $n$th ON during one channel coherence period can be rewritten as
\begin{equation}
	\begin{array}{lll}
{H_n}\left( f \right) &= \left| {{H_n}\left( f \right)} \right|{e^{j\angle {H_n}\left( f \right)}}\\ &= \sum\limits_{l = 0}^{L_n^c - 1} {\frac{{{G_T}{G_R}}}{{\sqrt {{L_{n,l}}\left( {{r_{n,p}},f} \right)} }}{\beta _{n,l}}\left( f \right){e^{ - j2\pi f{\tau _l}}}},\\
\end{array} 
\end{equation}
where $\left| {{H_n}\left( f \right)} \right|$ and ${e^{j\angle {H_n}\left( f \right)}}$ denote the amplitude and phase of the channel, respectively.
\vspace{-2mm}
\subsection{MIMO-OFDM UWA-ISAC Waveform   \label{sec-wav}}
\begin{comment}

\begin{figure}[t]
	\centering
	\includegraphics[width=0.5\textwidth]{plot.pdf}
	\caption{MIMO-OFDM ISAC transmit frame structure. }
	%————————————Fig.6————————————%
	\label{FigTradeoff_D_M}
\end{figure}

\end{comment}
The ISAC waveform transmitted by the SCN must simultaneously illuminate both ONs and targets under conditions of partial or complete absence of prior knowledge for spatial orientations. To synthesize an omnidirectional transmit beam pattern, the MIMO sonar requires $M_t$ mutually orthogonal waveforms allocated to its transmit array elements \cite{n9}. To exploit spatial diversity gain, block-type and interleaved-type OFDM signals in the frequency domain are widely adopted by MIMO radars and sonars \cite{n7,n9}. In comparison, the interleaved-type subcarrier structure enables more flexible waveform design for MIMO-OFDM ISAC systems. Notably, each pulse incorporates multiple zero-padded (ZP) OFDM symbols for the UWA-ISAC system. Each symbol serves dual functionalities: sensing and communication. Specifically, the first symbol is designed to broadcast resource allocation policies to downlink communication users. Subsequent symbols are exclusively allocated for arbitrary data transmission. For notational simplicity, the subsequent analysis in this paper focuses on a single data symbol to derive the system model. Nevertheless, the proposed framework inherently supports extensions to multi-symbol configurations. Thus, the baseband OFDM signal transmitted by the $m$th transmit array element can be expressed as
\begin{equation}
{{\mathbf{s}}_m} = {{\mathbf{F}}^H}\operatorname{diag} \left\{ {\mathbf{\bar p}} \right\}{{\mathbf{w}}_m}\operatorname{diag} \left\{ {\mathbf{\bar d}} \right\}, 
\end{equation}
where 
$\mathbf{F}\in {\mathbb{C}^{K \times K}}$ is the defined discrete Fourier transform (DFT) matrix with ${\left[ {\bf{F}} \right]_{q,k}} = {1 \mathord{\left/	{\vphantom {1 {\sqrt {{K}} }}} \right.	\kern-\nulldelimiterspace} {\sqrt {{K}} }}{e^{{{ - j2\pi kq} \mathord{\left/{\vphantom {{ - j2\pi nl} {{K}}}} \right.
\kern-\nulldelimiterspace} {{K}}}}}$, $K$ is the number of subcarrier in one OFDM symbol. Let $\mathcal{K} = \left[ {1,2, \cdots, K} \right]$ denotes the set of subcarrier index. ${\mathbf{\bar p}} = {\left[ {\sqrt {{p_1}} ,\sqrt {{p_2}} , \cdots ,\sqrt {{p_K}} } \right]^T}$ is the amplitude vector, ${\mathbf{p}} ={\left[ { {{p_1}} , {{p_2}} , \cdots , {{p_K}} } \right]^T}$ is the power allocation vector, and ${{\mathbf{w}}_m} \in {\left\{ {0,1} \right\}^{K \times 1}}$ is the subcarrier interleaved vector. It is noteworthy that the equal allocation of the entire $K$ subcarriers to the ${M_t}$ transmit array elements in this paper, which leads to the relationship
${\left\| {{{\mathbf{w}}_m}} \right\|_0} = {K \mathord{\left/{\vphantom {K {{M_t}}}} \right.\kern-\nulldelimiterspace} {{M_t}}}$. The vector ${\mathbf{\bar d}} = {{\mathbf{\bar d}}_1} + {{\mathbf{\bar d}}_2} +  \cdots  + {{\mathbf{\bar d}}_{{N_u}}}$ contains the modified data symbols sent to all ONs, where ${{\mathbf{\bar d}}_n \in \mathbb{C}^{K \times 1}}$ is the modified data symbols sent to the $n$th ON, and where the $k$th element is given by
\begin{equation}
{\bar d_{n,k}} = \left\{ {\begin{array}{*{20}{c}}
		{0{\text{   }},{\text{if }}{x_{n,k}} = 0} \\ 
		{{{\left( {{e^{j\angle {H_n}\left( {{f_k}} \right)}}} \right)}^H}{d_{n,v}},{\text{if }}{x_{n,k}} = 1} 
\end{array}} \right. ,
\end{equation}
where ${d_{n,v}}$ denotes the $v$th intended symbol of the $n$th ON, and it is from 
the pre-designed constellation dictionary
$\mathcal{J}=\{e^{2 \pi \cdot(0,1, \cdots, J-1) / J}\}$, where $J$ is the order of phase modulated. Let ${\mathbf{X}} = \left[ {{{\mathbf{x}}_1};{{\mathbf{x}}_2}; \cdots ;{{\mathbf{x}}_{{N_u}}}} \right]$ denote the frequency allocation matrix to all $N_u$ ONs. ${{\mathbf{x}}_n} = {\left[ {{x_{n,1}},{x_{n,2}}, \cdots ,{x_{n,K}}} \right]^T}$ is the frequency allocation vector, and it determines the subcarrier index of the data to be transmitted to the $n$th ON. Similarly, the $K$ subcarriers are equally allocated to $N_u$ ONs for downlink communication, i.e., ${\left\| {{{\mathbf{x}}_n}} \right\|_0} = {K \mathord{\left/
		{\vphantom {K {{N_u}}}} \right.
		\kern-\nulldelimiterspace} {{N_u}}}$.

Based on the above discussion, we can derive the transmit signal covariance matrix of the UWA-ISAC system
\begin{equation}
\begin{array}{lllll}
	{{\mathbf{R}}_{{\mathbf{SS}}}} &= \sum\limits_{{m_1} = 1}^{{M_t}} {\sum\limits_{{m_2} = 1}^{{M_t}} {{\mathbf{w}}_m^H} } {\mathbf{F}}\operatorname{diag} \left\{ {{\mathbf{p}}{{\mathbf{p}}^H}} \right\}\\
    & \quad\times \operatorname{diag} \left\{ {{\mathbf{\bar d}}{{{\mathbf{\bar d}}}^H}} \right\}{{\mathbf{F}}^H}{{\mathbf{w}}_m} \\ 
	&= {\left| {\operatorname{diag} \left\{ {\mathbf{\bar p}} \right\}{\mathbf{W}}} \right|^2} = \sum\limits_{k = 1}^K {p_k} {{\mathbf{W}\left[ k \right]}^{H}}{\mathbf{W}\left[ k \right]},
\end{array}
\end{equation}
where ${\mathbf{W}} = \left[ {{{\mathbf{w}}_1}; {{\mathbf{w}}_2}; \cdots ; {{\mathbf{w}}_{{M_t}}}} \right]$ is the subcarrier interleaved pattern consisting of the interleaved vectors of all the array elements. ${\mathbf{W}\left[ k \right]}$ denotes the $k$th row of matrix $\mathbf{W}$. For our proposed MIMO-OFDM UWA-ISAC system, each subcarrier is transmitted by only a single array element, so only one element in ${\mathbf{W}\left[ k \right]}, k \in \mathcal{K}$ is $1$ and the rest are $0$.
Further, the transmit beam pattern can be obtained
\begin{equation}
	\begin{array}{lll}
			G(\theta ) &= {{\mathbf{A}}^H}(\theta ){{\mathbf{R}}_{{\mathbf{SS}}}}{\mathbf{A}}(\theta )  \\ 
		&= \sum\limits_{k = 1}^{K} {p_k {{\mathbf{a}_t}^H}({f_k},\theta ){{\mathbf{W}\left[ k \right]}^{H}}{\mathbf{W}\left[ k \right]}{\mathbf{a}_t}({f_k},\theta )} = {{P_{{\text{total}}}}},\\ 
	\end{array}
	\label{eq:eq10}
\end{equation}
where ${{\mathbf{a}}_t}({f_k},\theta ) = \left[ {1,e^{ {j{f_k}\eta \left( \theta  \right)} }, \cdots ,} \right.{\left. {e^{ {j\left( {{M_t} - 1} \right){f_k}\eta \left( \theta  \right)} }} \right]^T}$ is the transmit steering vector corresponding to the $k$th subcarrier, $\eta \left( \theta  \right) = 2\pi {d_t}\sin \theta /c$,
$d_t$ is the spacing between adjacent transmit elements, and it is an integer multiple of $d$.
$f_k={f_l} + (k-1) \Delta f $ is the frequency of the $k$th subcarrier, ${f_l}$ is the carrier frequency, and $\Delta f $ is the subcarrier spacing. Therefore, the bandwidth of MIMO-OFDM signal is $B=K\Delta f$. Moreover, ${{P_{{\text{total}}}}}$ is the total transmit power.
Observing (\ref{eq:eq10}), we find that the radiated power is constantly equal to the total power ${{P_{{\text{total}}}}}$ at any spatial angle. Consequently, the UWA-ISAC system can establish communication links and sense targets simultaneously in different directions.

%MIMO-OFDM ISAC system的时域发射波形表示为

\section{MIMO-OFDM UWA-ISAC System Performance Analysis and Resource Optimization}

\subsection{System Performance Analysis}
%In this section, we evaluate the comprehensive performance of the MIMO-OFDM UWA-ISAC system in three dimensions: transmission, communication, and sensing, and establish a resource optimization model based on the analysis results. Finally, we solve the non-convex and complex optimization problem by using a faster converging grouped randomized search (GRS) scheme.
\subsubsection {Transmission Performance}
 In the proposed MIMO-OFDM ISAC system, the passband signal at the $m$th transmit element is formulated as
\begin{equation}
	\begin{array}{ll}
		{s_m}\left( t \right) = &\operatorname{Re} \left[ {\sum\limits_{k = 0}^{K - 1} {\sqrt {{p_k}} {w_{m,k}}\bar d\left[ k \right]}  \cdot {e^{j2\pi {f_k}t}}g  \left( t \right)} \right],\\
	\end{array}
\end{equation}
where $t \in \left[ {0,T_p} \right]$, $T_p=T+T_g$ is the duration of one pulse, $T$ and $T_g$ denote the OFDM symbol length and the ZP duration between symbols, respectively. The gap can avoid inter-symbol interference for communication and sensing. $g  \left( t \right)$
is the filter response function of the transmitter.
%but also increases the maximum unambiguous distance for sensing

The peak-to-average power ratio (PAPR) serves as a critical metric for evaluating the transmission performance of, as it indirectly impacts both ISAC capabilities. Due to the limited dynamic range of the power amplifier, excessively high PAPR induces nonlinear distortion in the amplifier, thereby degrading the effective transmit power. Consequently, to maximize the power efficiency of the MIMO-OFDM UWA-ISAC system, the transmit signal at each antenna element is required to exhibit a low PAPR, which enhances communication and sensing performance. 
The PAPR of the transmit signal at the $m$th element is defined as
\begin{equation}
{\text{PAPR}}\left\{ {{s_m}} \right\} = 10\log \left( {\frac{{\mathop {\max }\limits_t \left\{ {{{\left| {{s_m}\left( t \right)} \right|}^2}} \right\}}}{{\operatorname{E} \left\{ {{{\left| {{s_m}\left( t \right)} \right|}^2}} \right\}}}} \right),
\end{equation}
where $t \in \left( {0,T} \right]$. Given that $\bar d$ has unit amplitude, the average transmit power per array element for a specified number of subcarriers is related to ${p_k}$, while the peak power depends not only on ${p_k}$ but also on the subcarrier interleaved pattern ${\mathbf{W}}$ and subcarrier allocation scheme ${\mathbf{X}}$.

\subsubsection {Communication Performance \label{sec-com}}

Since the SCN utilizes different frequency resources to simultaneously transmit information to multiple communication users, the co-channel interference can be neglected. Therefore, based on the channel model and the UWA-ISAC waveform in Section \ref{sec-wav},  the achievable communication rate of the $n$th ON connected with the SCN can be calculated as
\begin{equation}
{R_n} = \sum\limits_{k = 1}^K {{x_{n,k}}\Delta f{{\log }_2}\left( {1 + \frac{{{p_k}{{\left| {{H_n}\left( {{f_k}} \right)} \right|}^2}}}{{\sigma _k^2}}} \right)},
\end{equation}
where ${\sigma _k^2}$ is noise variance corresponding to the $k$th subcarrier, which is a function of frequency and can be calculated from turbulence, shipping, waves, and thermal source of noise \cite{b15}. 
The achievable communication rate is commonly adopted as the objective function for resource allocation in existing literature, particularly in MIMO-OFDM and orthogonal frequency division multiple access systems. Since sensing functionality is not considered in these scenarios, differentiated power allocation can effectively balance multiple communication receivers. However, for our proposed UWA-ISAC system, larger power disparities among subcarriers will lead to more significant degradation in sensing performance, as will be elaborated in Section \ref{sec-sen}. Therefore, to better serve multiple distributed communication users while preventing frequency resources from being disproportionately allocated to distant users, this paper employs the product of communication rate and range (PRR) as the communication performance metric. 
Hence, the overall achievable product of communication rate and range for the SCN can be expressed as
\begin{equation}
	PRR = \sum\limits_{n = 1}^{N_u} {{PRR_n}}=\sum\limits_{n = 1}^{N_u} {{R_n}}r_n,
			\label{eq:eq16}
\end{equation}	
where $r_n$ is the straight line distance between the SCN and the $n$th ON. We can find from (\ref{eq:eq16}) that the achievable communication rate is only related to the frequency and power allocation scheme when the distances are almost constant.

\subsubsection {Sensing Performance\label{sec-sen}}
Target echoes carry both range and angular information, the received signal at the $k$th subcarrier of the MIMO sonar is expressed as
\vspace{-2mm}
\begin{equation}
	\begin{array}{llll}
	{{\mathbf{y}}^{\text{s}}}_k\left( t \right) =& \sum\limits_{q = 1}^Q {{\gamma _q}{e^{ - j2\pi {f_c}\tau _q^{\text{s}}}}{{\mathbf{a}}_r}\left( {{f_k},{\theta _{rq}}} \right){\mathbf{a}}_t^T\left( {{f_k},{\theta _{tq}}} \right)}  \\ 
	&\times {{\mathbf{s}}_k}\left( {t - \tau _q^{\text{s}}} \right) + {\mathbf{v}}_k^{\text{s}}\left( t \right), \\ 
	\end{array}
\label{eq:eq17}
\end{equation}
where ${\gamma _q}$ and $ \tau_q^{\text{s}}$ are the scattering coefficient and the time delay of $q$th target, respectively, ${\mathbf{v}}_k^{\text{s}}\left( t \right) = {\left[ {v_{1,k}^{\text{s}}\left( t \right),v_{2,k}^{\text{s}}\left( t \right), \cdots ,v_{{M_r},k}^{\text{s}}\left( t \right)} \right]^T}$ is the noise vector received by the MIMO sonar, and  ${{\mathbf{a}}_r}\left( {{f_k},{\theta}} \right)$ is receive steering vector. 
\begin{comment}
, and it can be expressed as 
\begin{equation}
	\begin{array}{ll}
		{\mathbf{a}_r}({f_k},\theta ) =& \left[ {1,\exp \left( {j2\pi {f_k}d_r\sin \theta /c} \right), \cdots ,} \right.\\ &{\left. { \exp \left( {j\left( {{M_r} - 1} \right)2\pi {f_k}d_r\sin \theta /c} \right)} \right]^T}\\
	\end{array},
	\label{eq:eq11}
\end{equation}
where $d_r$ is the spacing between adjacent receive elements, again an integer multiple of $d$. Generally, setting $d_t = M_r d_r$ can maximize the spatial resolution. 
\end{comment}
Next, matched filtering and and beamforming are performed on the received signal to obtain a joint spectrum in the space-time domain
\begin{equation}
	\begin{array}{llll}
	{{\mathbf{C}}_k}\left( {\tau ,\theta } \right) 
	&= \sum\limits_{q = 1}^Q {{p_k}{\gamma _q}{e^{ - j2\pi {f_k}\tau _q^s}}{\mathbf{A}}_{tr}^H\left( {{f_k},\theta } \right)}  \\ 
	&\quad \times {{\mathbf{A}}_{tr}}\left( {{f_k},{\theta _{rq}}} \right) {\text{diag}}\left\{ {{r_k}\left( {\tau } \right)} \right\} + {\mathbf{v'}}\left( \tau  \right),
	\end{array}  
\end{equation}
where ${{\mathbf{r}}_k}\left( {\tau } \right)$ denotes the normalized auto-correlation function of ${\mathbf{s}}_k$, ${{\mathbf{A}}_{tr}}\left( {{f_k},{\theta _{rq}}} \right) = {{\mathbf{a}}_r}\left( {{f_k},{\theta _{rq}}} \right){\mathbf{a}}_t^T\left( {{f_k},{\theta _{tq}}} \right)$, and ${\mathbf{v'}}\left( \tau  \right) = \int_T^{{T_p}} {{{\mathbf{v}}_k}\left( t \right){\mathbf{s}}_k^H\left( {t - \tau } \right)} dt$ is the processed noise vector.  
The spatial resolution of the joint spectrum depends on the equivalent aperture of the MIMO sonar, and the delay resolution is determined by the transmit signal bandwidth of the MIMO-OFDM UWA-ISAC system. Therefore, we focus more on the delay profile of the joint spectrum. We extract the time-delay profile of the $q$th target, denoted as
\vspace{-2mm}
\begin{equation}
c_q\left( \tau  \right) = {\gamma _q}\sum\limits_{k = 0}^{K - 1} {{p_k}{e^{j2\pi {f_k}\left( {\tau  - \tau _q^s} \right)}}}  + z_q \left( \tau  \right) + {\mathbf{v''}}_q \left( \tau  \right), 
\label{eq:eq20}
\end{equation}
where $z_q\left( \tau  \right)$ denotes the sum of the beamforming output components of the other $Q-1$ targets, and  ${\mathbf{v''}}_q \left( \tau  \right)$ denotes the noise after beamforming processing. Equation (\ref{eq:eq20}) demonstrates that the beam-domain delay profile of MIMO-OFDM UWA-ISAC is identical to that of conventional OFDM. The first term on the right-hand side of (\ref{eq:eq20}) represents a modified auto-correlation function of the equivalent OFDM signal. Note that minimizing the variance of $\left\{ {{p_k}} \right\}_{k = 1}^K$ achieves the lowest delay sidelobes, which is critical for ensuring superior sensing performance \cite{n2}.

\subsection{Resource Optimization Model for UWA-ISAC}
We consider an underwater scenario requiring high sensing performance (e.g., multiple threatening targets) while concurrently maximizing multi-user communication capacity. The joint transmission, communication, and sensing performance primarily depend on the allocated power and frequency resources to each transmit array element and communication user, i.e., power allocation vector ${\mathbf{p}}$, subcarrier interleaved
pattern ${\mathbf{W}}$, and frequency allocation matrix ${\mathbf{X}}$. Consequently, we formulate the optimization objective to maximize the total achievable communication PRR, subject to guaranteed minimum ranging sidelobe levels and acceptable PAPR. Based on this framework, the subcarrier interleaved and resource (power and frequency) allocation problem for the MIMO-OFDM UWA-ISAC system is expressed as
\vspace{-3mm}
\begin{comment}
\begin{equation}
\begin{gathered}
		\begin{array}{*{20}{l}} \hfill
		\mathop {\max }\limits_{{\mathbf{W}},{\mathbf{X}}} & R  \\
		{s.t.}&{\forall m \in {\mathcal{M}_t},{{\left\| {{{\mathbf{w}}_m}} \right\|}_1} = \frac{K}{{{M_t}}}}&{{C_1}} \\ 
		{}&{\forall m \in {\mathcal{M}_t},\forall k \in \mathcal{K},{w_{m,k}} \in \left\{ {0,1} \right\}}&{{C_2}} \\ 
		{}&{\forall n \in {\mathcal{N}_u},{{\left\| {{{\mathbf{x}}_n}} \right\|}_1} = \frac{K}{{{N_u}}}}&{{C_3}} \\ 
		{}&{\forall n \in {\mathcal{N}_u},\forall k \in \mathcal{K},{x_{n,k}} \in \left\{ {0,1} \right\}}&{{C_4}} \\ 
		{}&{\forall n \in {\mathcal{N}_u},{R_n} \geqslant {R_{\min }}}&{{C_5}} \\ 
		{}&{\forall m \in {\mathcal{M}_t}, {\text{PAPR}}\left\{ {{s_m}} \right\} \leqslant \text{PAPR}_0}&{{C_6}} \\ 
		{}&{\forall k \in {\mathcal{K}},{p_k} = \frac{{{P_{{\text{total}}}}}}{K}}&{{C_7}} \\ 
		{}&{\forall k \in {\mathcal{K}},\sum\limits_{\forall m \in {\mathcal{M}_t}} {{w_{m,k}}}  = 1,\sum\limits_{\forall n \in {\mathcal{N}_u}} {{x_{n,k}}}  = 1}&{{C_8}} 
	\end{array} \hfill \\ 
\end{gathered}
	\label{eq:eqP}
\end{equation}
\end{comment}
\begin{comment}

\begin{equation}
\max_{\mathbf{W}, \mathbf{X}} \quad PRR
\tag{21}\label{eq:21}
\end{equation}
\end{comment}
\begin{subequations}\label{eq:eqP}
\begin{align}
\max_{\mathbf{W}, \mathbf{X}} \quad & PRR
\tag{16}\label{eq:21} \\
\text{s.t.}\quad & \forall m \in \mathcal{M}_t,\quad \|\mathbf{w}_m\|_1 = \frac{K}{M_t}
\tag{16a}\label{eq:21a} \\
& \forall m \in \mathcal{M}_t,\forall k \in \mathcal{K},\quad w_{m,k} \in \{0,1\}
\tag{16b} \label{eq:21b}\\
& \forall n \in \mathcal{N}_u,\quad \|\mathbf{x}_n\|_1 = \frac{K}{N_u}
\tag{16c} \label{eq:21c}\\
& \forall n \in \mathcal{N}_u,\forall k \in \mathcal{K},\quad x_{n,k} \in \{0,1\}
\tag{16d} \label{eq:21d}\\
& \forall n \in \mathcal{N}_u,\quad PRR_n \ge PRR_{\min}
\tag{16e} \label{eq:21e}\\
& \forall m \in \mathcal{M}_t,\quad \mathrm{PAPR}\{s_m\} \le \mathrm{PAPR}_0
\tag{16f} \label{eq:21f}\\
& \forall k \in \mathcal{K},\quad p_k = \frac{P_{\mathrm{total}}}{K}
\tag{16g} \label{eq:21g}\\
& \forall k \in \mathcal{K},\quad \sum_{m \in \mathcal{M}_t} w_{m,k} = 1,\quad \sum_{n \in \mathcal{N}_u} x_{n,k} = 1.
\tag{16h}\label{eq:21h}
\end{align}
\end{subequations}
 Constraints  ~\eqref{eq:21a} and ~\eqref{eq:21b} enforce uniform subcarrier allocation across transmit array elements, constraints ~\eqref{eq:21c} and ~\eqref{eq:21d} impose identical bandwidth allocation for multi-user data transmission, constraints ~\eqref{eq:21e} and ~\eqref{eq:21f} specify the minimum PRR requirements for communication users and PAPR constraints for transmit signals, respectively. Furthermore, constraint ~\eqref{eq:21g} ensures uniform power distribution across subcarriers to preserve sensing performance, and constraint ~\eqref{eq:21h} guarantees that each subcarrier can only be assigned to one transmit element and one communication user.

\subsection{TDGRS scheme for solving (\ref{eq:eqP})}
The optimization in \eqref{eq:eqP} is inherently a two-dimensional mixed-integer nonconvex problem, which poses significant challenges in finding the optimal solution. Unlike previous solution methods\cite{n7,b15}, we develop a TDGRS strategy to quickly obtain the locally optimal subcarrier's interleaved pattern and allocation scheme. By filtering out infeasible solutions and selecting the most optimal configurations, the TDGRS aims to maximize the system PRR while satisfying the constraints set by the UWA environment. As shown in \textbf{Algorithm 1}, TDGRS follows a structured approach, outlined in the steps below: 
\textbf{I. Group-wise Optimization and Reassignment}: The algorithm proceeds by sequentially performing randomized reassignments for each group. For each group $g$, random permutations of the interleaved pattern $\mathbf{W}_0^g$ and subcarrier allocation $\mathbf{X}_0^g$ are performed. In this phase, subcarrier permutations are conducted for each group independently, allowing the system to explore different configurations. These permutations are then evaluated based on their feasibility, considering the constraints of PRR and PAPR. \textbf{II. Feasibility Check and Solution Selection}: After each group-wise reallocation, the total PRR is computed. For each feasible configuration that satisfies the system constraints, the solution with the highest PRR is retained. The optimal interleaved subcarrier pattern $\mathbf{W}^*$ and allocation matrix $\mathbf{X}^*$ are updated when a better solution is found, ensuring the system achieves maximum PRR while satisfying the constraints. 
%Specifically, the overall framework of the TDGRS scheme is summarized in \textbf{Algorithm 1}.

\begin{algorithm}[h]
	\caption{TDGRS Algorithm for Joint Optimization of Subcarrier Interleaved and Allocation}
	\label{alg:GRS}
	\begin{algorithmic}
		\renewcommand{\algorithmicrequire}{\textbf{Input:}}
		\renewcommand{\algorithmicensure}{\textbf{Output:}}
		\REQUIRE Parameters $K$, $M_t$, $N_u$ and  $\{d_{n,v}\}_{n=0}^{N_u}$, channel matrices $\{\mathbf{H}_n\}_{n=0}^{N_u}$ and algorithm parameters $G$, $E_1$, $E_2$.
		\ENSURE Optimized $\mathbf{W}^*$, $\mathbf{X}^*$, and $PRR$.
		%\vspace{1ex}
		\STATE \textbf{Initialize}: Generate $\mathbf{W}_0$, $\mathbf{X}_0$ via sequential interleaved scheme and partition $K$ subcarriers into $G$ groups, and set $\Delta \leftarrow 0$ and $(\mathbf{W}^*, \mathbf{X}^*) \leftarrow \varnothing$.
		%\vspace{1ex}
		\FOR{$g = 1$ to $G$}
		\STATE \textbf{Step I: Filtering of subcarrier allocation }
		\FOR{$e_1 = 1$ to $E_1$}
		\STATE Randomly reassign $\mathbf{X}_0^g$ to $\mathbf{X}_{e_1}^g$ and compute $PRR_n$.
		\ENDFOR  
		\STATE Filter out solutions that satisfy constraint~\eqref{eq:21e}, and collect feasible $\mathbf{X}_{e_1}^g$ into $\mathcal{X}_{\text{feasible}}^g$.
		%\vspace{1ex}
		\STATE \textbf{Step II: Joint optimization of subcarrier interleaved and allocation}
		\FOR{$e_2 = 1$ to $E_2$}
		\STATE Randomly reassign $\mathbf{W}_0^g$ to $\mathbf{W}_{e_2}^g$ and combine it with each element in $\mathcal{X}_{\text{feasible}}^g$, and compute PAPRs of all OFDM signals.
		\ENDFOR
		\STATE Filter out solutions that satisfy~\eqref{eq:21f}, and compute ~\eqref{eq:21} and the corresponding $\mathbf{W}^g$ and $\mathbf{X}^g$.
		\STATE If $ PRR > \Delta$, update $\Delta \leftarrow PRR$, $(\mathbf{W}^*, \mathbf{X}^*) \leftarrow ( \mathbf{W}^g, \mathbf{X}^g) $ and $(\mathbf{W}_0, \mathbf{X}_0) \leftarrow (\mathbf{W}^g, \mathbf{X}^g)$.
		\ENDFOR
		%\vspace{1ex}
		%\STATE \textbf{Output}: $\mathbf{W}^*$, $\mathbf{X}^*$, and $PRR$.
	\end{algorithmic}
\end{algorithm}

\section{Simulation Experiment Results}
    
\begin{figure}[t]
	\centering
	\includegraphics[width=0.98\textwidth]{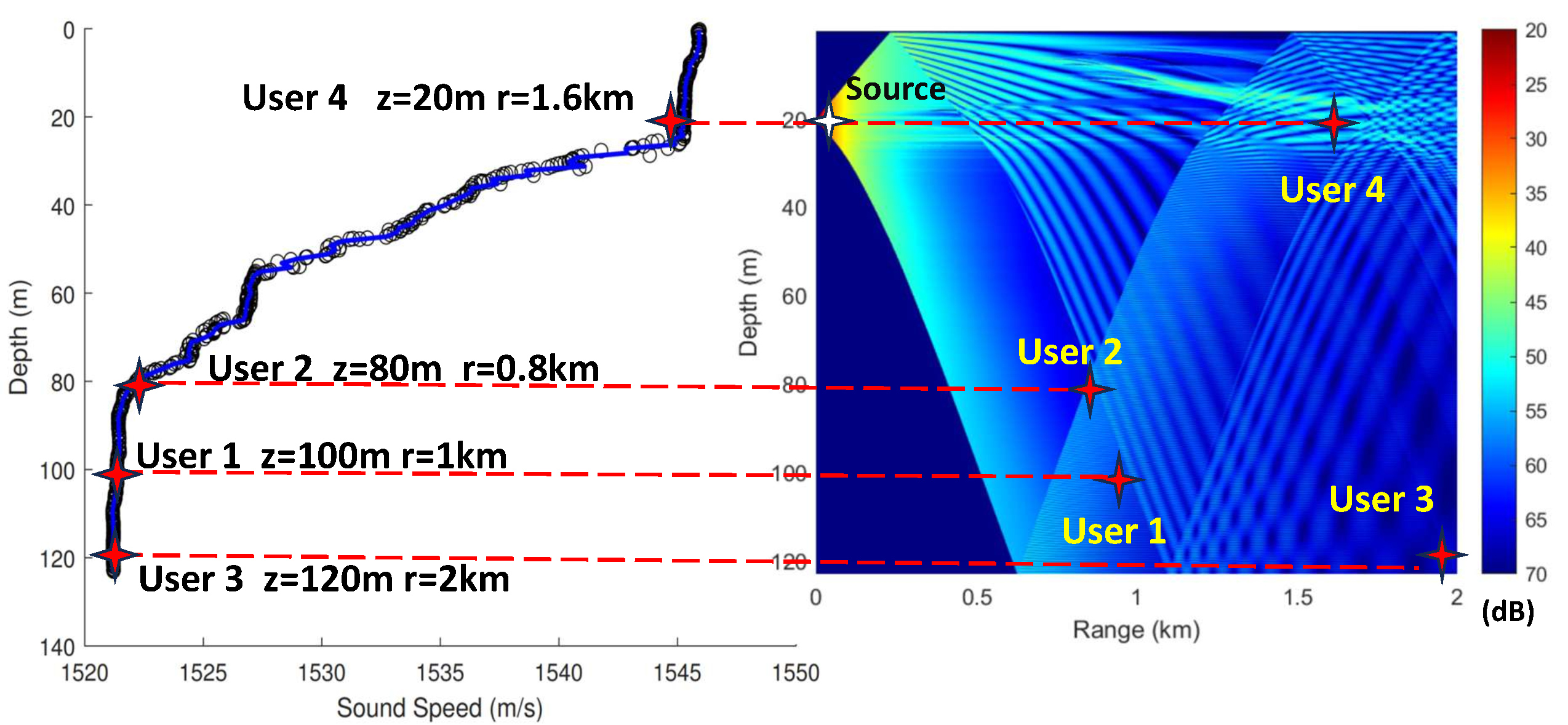}
	\caption{Real-world UWA environment and experiment layout. }
	%————————————Fig.2————————————%
	\label{Figenv}
\end{figure}

This section evaluates the effectiveness of our proposed UWA-ISAC  resource optimization framework under real-world UWA parameters measured in the South China Sea. All nodes are located in a shallow sea with a depth of $120$ m.  
The SCN carries a horizontal sonar array of 8 array elements placed at a depth of $20$ m. The main system parameters are set as follows unless explicitly stated otherwise. Total transmit  power of  SCN is $1$ W. The carrier frequency is \(f_l = 1 \text{ kHz}\), with a bandwidth of \(B = 4 \text{ kHz}\), and number of subcarriers is $K = 1024$. The results were obtained through 2000 Monte Carlo experiments. The SCN serves $N_u = 4$ ONs, comprising three bottom nodes and one near-surface node. The four users are placed at specific depths and horizontal ranges, with the exact locations indicated in Fig. \ref{Figenv}. The underwater ray tracing is implemented by Bellhop to calculate and analyze the UWA environment. Due to the large difference in the sound field structure for different users, there is a large optimization space for the subcarrier allocation scheme.

\begin{figure}[t]
	\centering
	\begin{minipage}[b]{0.45\linewidth}
		\centering
		\includegraphics[width=\linewidth]{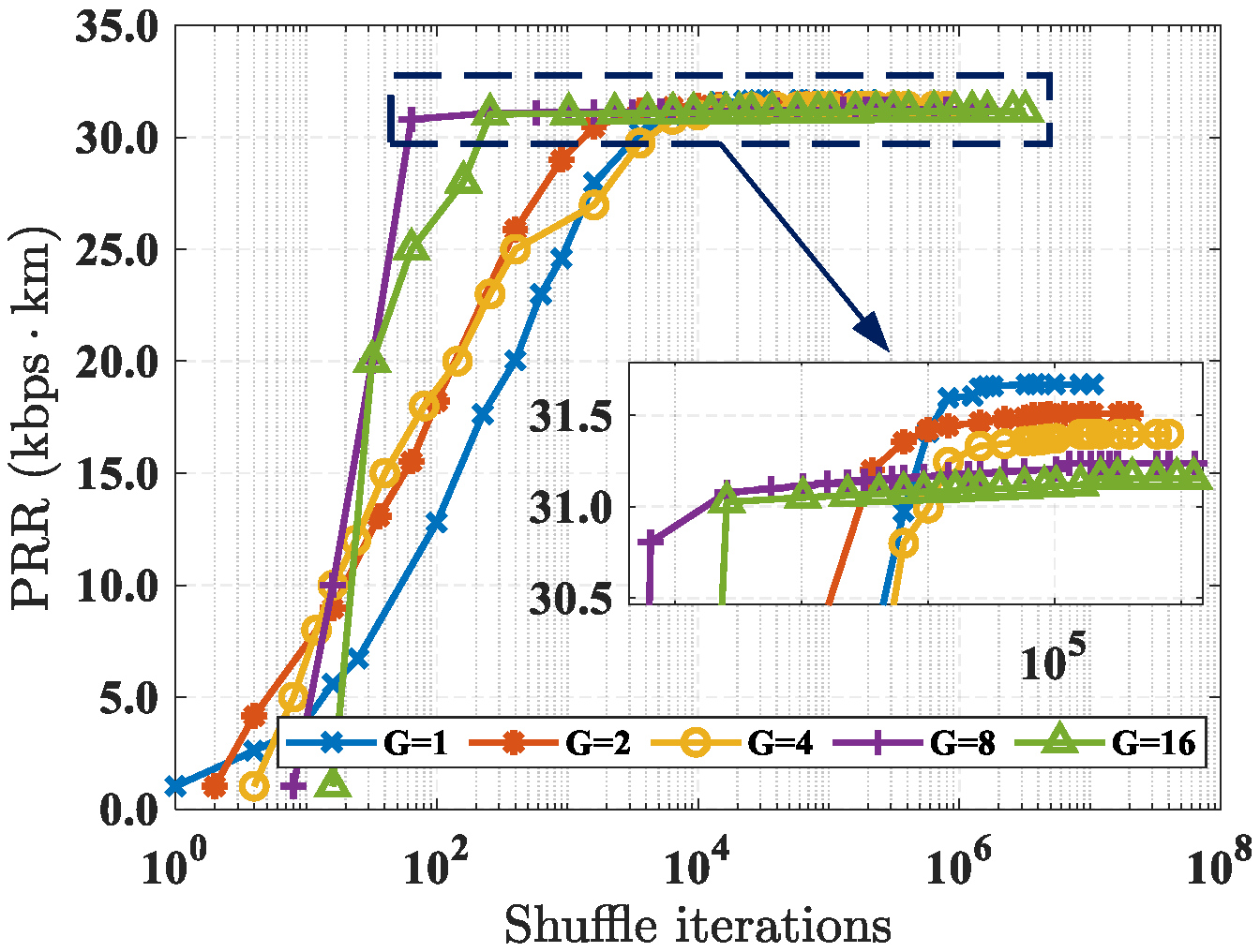}
		\caption*{(a) Different $G$.} % 使用*避免编号
		\label{sG}
	\end{minipage}
	\hfill
	\begin{minipage}[b]{0.45\linewidth}
		\centering
		\includegraphics[width=\linewidth]{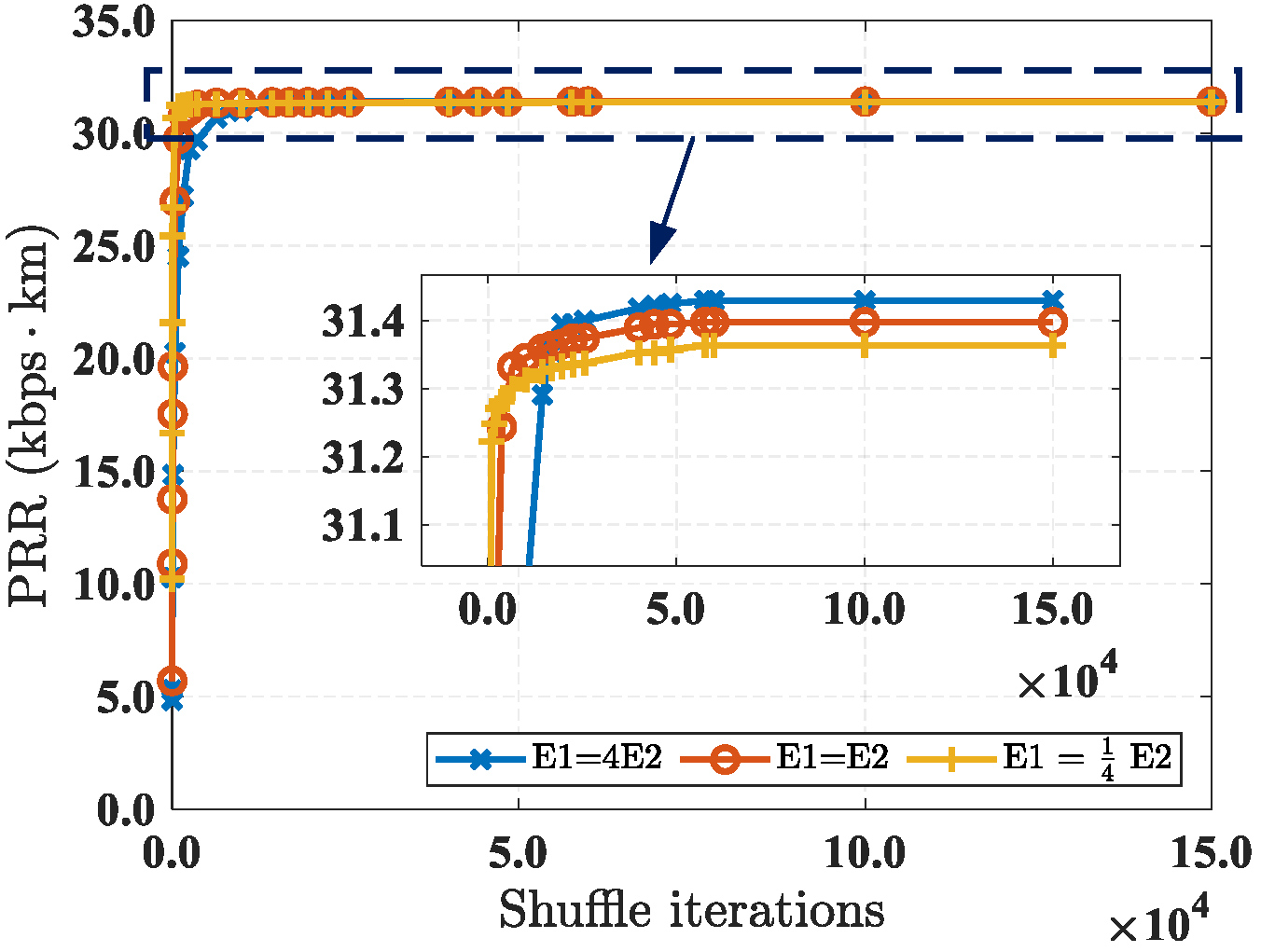}
		\caption*{(b) Different $E_1$ and $E_2$.} % 使用*避免编号
	\end{minipage}
	\caption{Convergence analysis of TDGRS algorithm.} % 整体标题
	\label{re1}
\end{figure}

Fig. \ref{re1} presents the multi-user communication performance as a function of shuffle iterations for various grouping configurations. Observation of Fig. \ref{re1} (a) reveals the PRR  as the number of shuffles increases. Notably, the higher the number of groups $G$, the faster the convergence since smaller group sizes facilitate the faster finding of locally optimal solutions. 
However, regarding ultimate convergence performance, fewer groups are preferable, as they yield solutions closer to the global optimum. In particular, the TDGRS (G=8) requires only 64 shuffle iterations to converge, which is merely 10\% of that required by conventional exhaustive search scheme (G=1), at the cost of a marginal $0.5$ kbps $\cdot$ km degradation in PRR. This demonstrates that the proposed method achieves significant computational efficiency gains with negligible performance compromise. Fig. \ref{re1} (b) shows that a larger $E_1$ achieves a superior PRR under deterministic total shuffle iterations, while a larger $E_2$ facilitates fast convergence.

\begin{figure}[t]
    \centering
    \begin{minipage}[b]{0.46\linewidth}
        \centering
        \includegraphics[width=\linewidth]{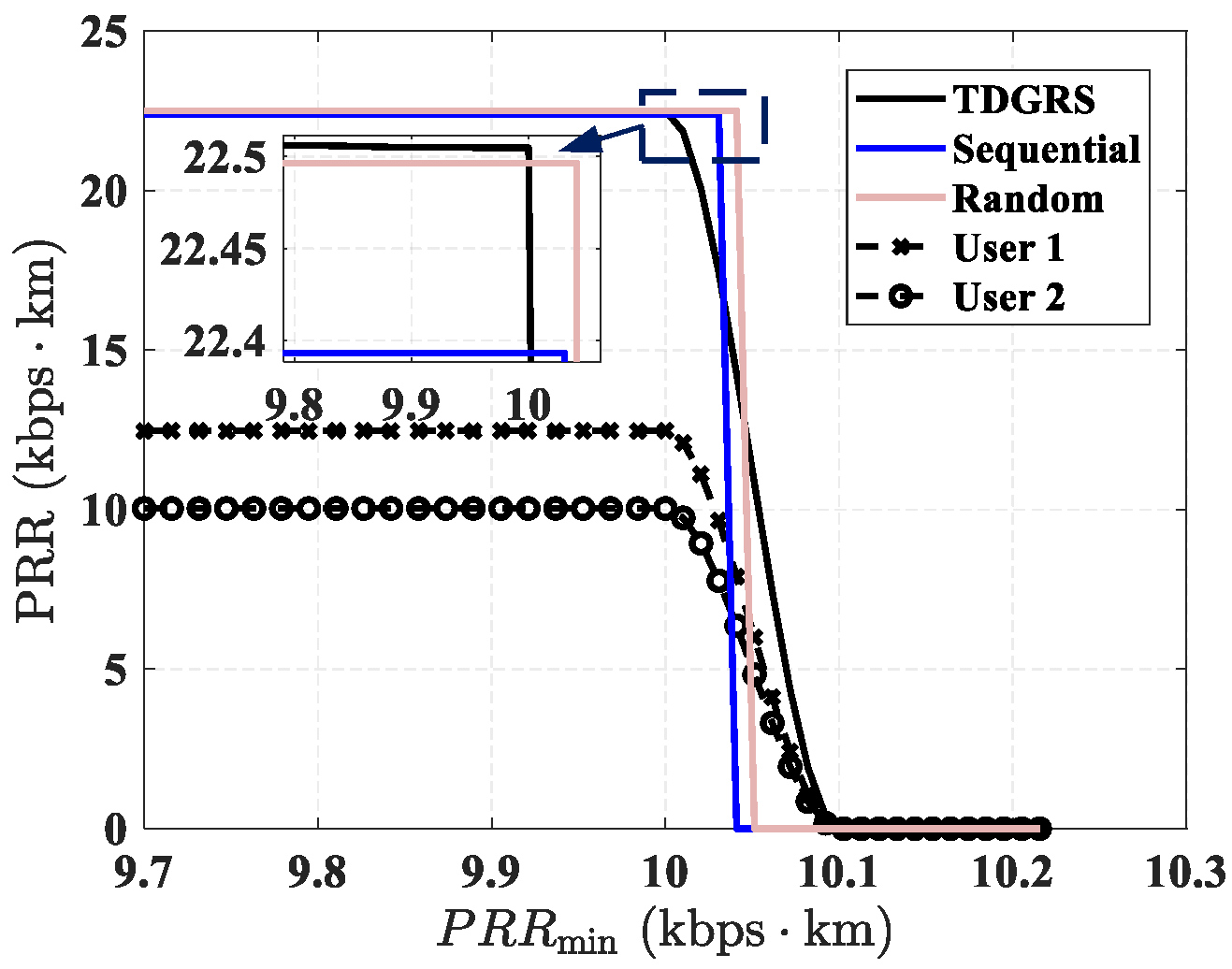}
        \caption*{(a) $N_u=2$.} % 使用*避免编号
    \end{minipage}
    \hfill
    \begin{minipage}[b]{0.46\linewidth}
        \centering
        \includegraphics[width=\linewidth]{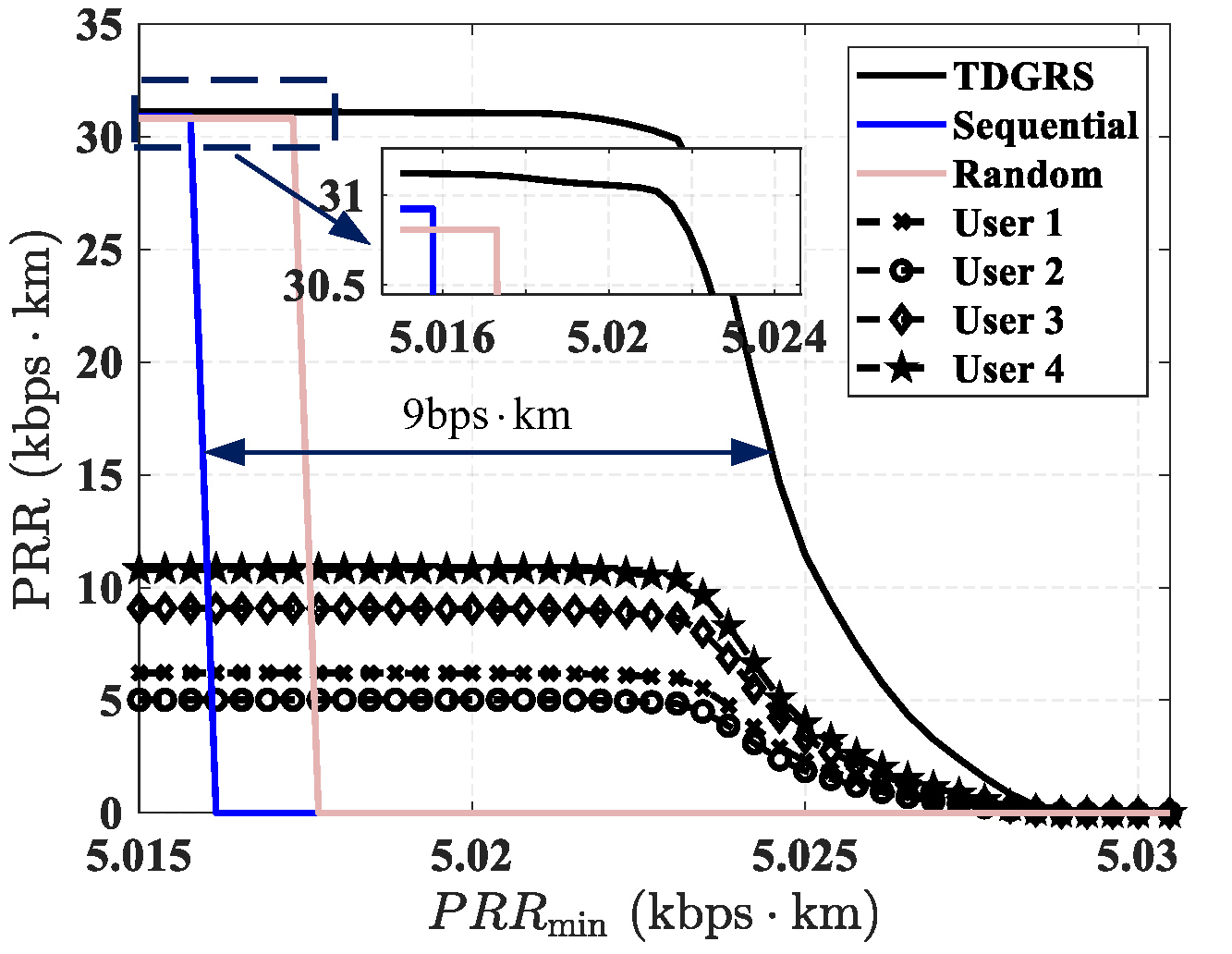}
        \caption*{(b) $N_u=4$.} % 使用*避免编号
    \end{minipage}
    \caption{PRR versus $PRR_{min}$ with $\mathrm{PAPR}_0=8.5$ dB.} % 整体标题
    \label{re2}
\end{figure}

 Fig. \ref{re2} presents PRR with respect to $PRR_{\min}$ under different $N_u$. 
 In general, PRR decreases as the threshold increases for both cases. This decline is primarily due to the reduced number of feasible configurations that satisfy the threshold, eventually leading to no feasible solution. The overall reduction in capacity is often attributed to the “bottleneck effect” caused by the worst-performing user (e.g., User 2), whose communication quality becomes a limiting factor for the system. As the rate threshold increases, the system must reallocate subcarriers among users to satisfy the constraints, decreasing the optimal PRR. However, our TDGRS (G=8) demonstrates superior robustness compared to sequential and random schemes. This is because TDGRS dynamically reallocates subcarriers as the threshold changes, allowing PRR to decline gradually. In contrast, two baseline schemes do not always yield feasible solutions. As shown in Fig. \ref{re2} (b), their acceptable PRR thresholds are more than $9$ bps$\cdot$km lower than TDGRS in the four-user case.

\begin{comment}
\begin{figure}[t]
	\centering
	\includegraphics[width=0.4\textwidth]{capacityuser.eps}
	\caption{Variation of the optimal rate with respect to the single-user communication rate threshold (constraint \eqref{eq:21e}). }
	%————————————Fig.5————————————%
	\label{re2}
\end{figure}
\end{comment}

\begin{figure}[t]
	\centering
	\includegraphics[width=0.8\textwidth]{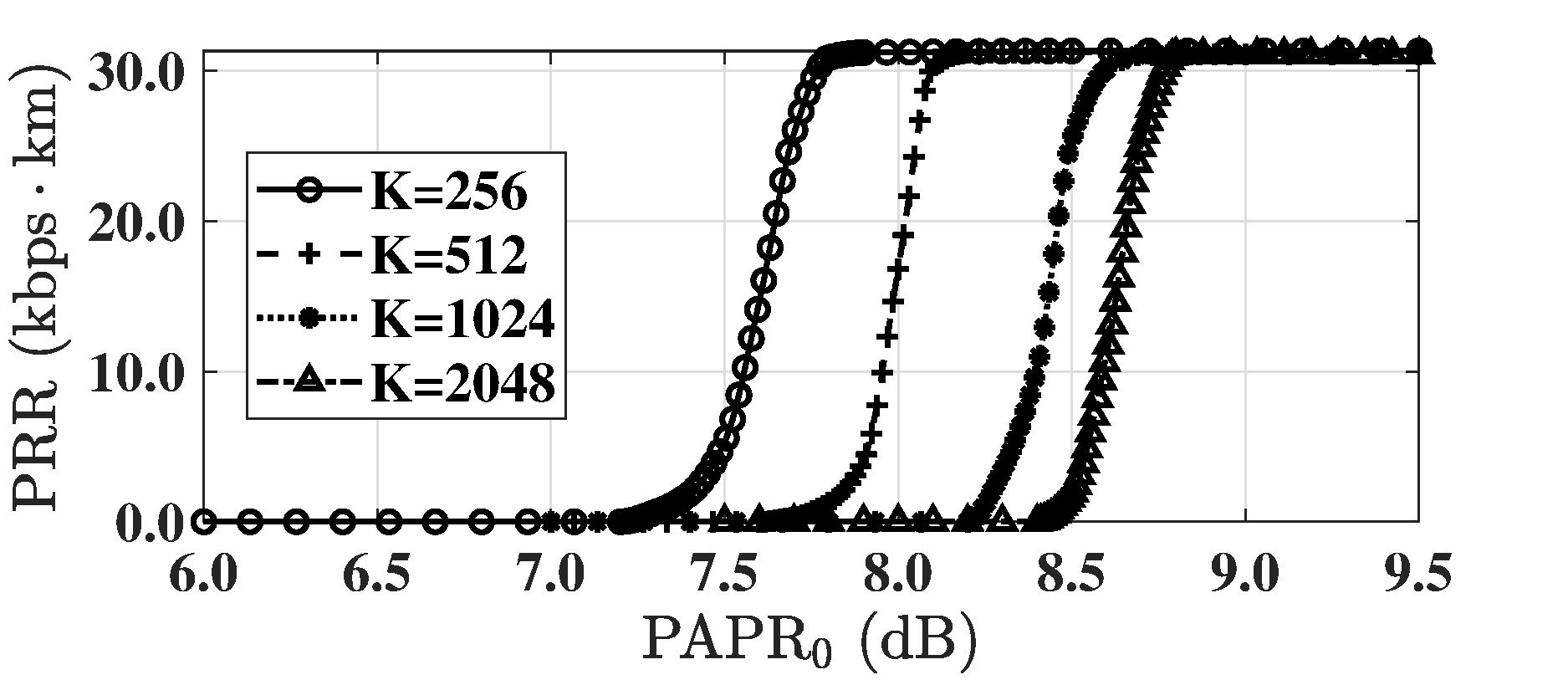}
	\caption{PRR versus $\mathrm{PAPR}_0$ with $PRR_{min}=4$ kbps$\cdot$km. }
	%————————————Fig.6————————————%
	\label{re3}
\end{figure}

Furthermore, the PRR as a function of the PAPR threshold for each transmit element is shown in Fig. \ref{re3}. We evaluate the relationship between PAPR and PRR under different numbers of subcarriers and find that constraint \eqref{eq:21f} also significantly affects the PRR of the UWA-ISAC system. This is because when the data to be transmitted to each user is determined, $\mathbf{W}$ directly controls the PAPR of OFDM signals emitted by each antenna element, which in turn affects $\mathbf{X}$. The PAPR constraint may prevent frequency resources from being allocated according to the optimal PRR. As the PAPR threshold relaxes, more feasible solutions become available, thereby optimizing the system’s PRR. Notably, although Fig. \ref{re3} shows that a looser PAPR constraint improves PRR, its unrestricted increase will reduce transmission efficiency, ultimately degrading the sensing and communication performance. Consequently, practical implementations should establish appropriate thresholds by leveraging the quantitative relationship between PAPR and the transmit source level to balance performance tradeoffs.

\section{Conclusion}

This paper has proposed an interleaved OFDM-based UWA-ISAC framework to address the challenges of simultaneous multi-user communication and multi-target sensing in IoUT. Moreover, we have evaluated the UWA-ISAC system from three dimensions: transmission, communication, and sensing, thereby establishing a resource optimization problem that maximizes PRR while ensuring transmission and sensing performance. Furthermore, to solve this non-convex complex problem, we have developed a TDGRS algorithm to obtain optimal local solutions quickly. The simulation results have demonstrated the superiority of the proposed UWA-ISAC system and resource optimization scheme. Future work will focus on dynamic resource allocation, considering sensing and communication tradeoffs.

\section*{Acknowledgment}

This work was supported in part by the National Natural Science Foundation of China under Grant 62401317 and 623B2060, in part by the Science and Technology on Sonar Laboratory Project under Grant JCKY2024207CE01. 
\begin{comment}

\newpage
\end{comment}
\bibliographystyle{IEEEtran}
\bibliography{reference}

\end{document}